\documentstyle[11pt]{article}
\date{}
\begin{document}
\sloppy
\title{Can the energy density of gravitational field be interpreted as dark energy?}
\author{V. Majern\'{\i}k\\
Institute of Mathematics Slovak Academy of Sciences,\\ SK-814 73
Bratislava, \v Stef\'anikova 49, Slovak Republic}
\maketitle

\begin{abstract} After a brief
review of the Maxwell-like approach to gravity we consider the issue
of the negative energy of gravitational field which is a consequence
of the field approach to the phenomenon of gravitation. Due to the
existence of the negative field energy {\it within} a mass body its
total energy content is smaller than the positive energy assigned to
its mass energy. We study the total energy content of a spherically
symmetrical mass body having constant matter density, and show that
its total energy content depends on its radius. We show that under
certain circumstances, the total energy content of a mass body
achieves negative values so that the force at its surface becomes
repulsive. We apply this idea to the evolution of universe filled by
matter and the negative energy density of its gravitational field.
Since the negative energy density causes the negative pressure it
might be considered as an agent which causes the acceleration of the
universe.
\end{abstract}

\section{Introduction}
An interesting development seems to take place in cosmology during
the last few years. The evidence continues to mount that the
expansion of the universe is accelerating  rather than slowing down.
Several astrophysical groups (Tonry et al. 2003 \cite{TO}; Barris et
al. 2004 \cite{BAA}; Riess et al. 2004 \cite{RS}) have recently
updated the original supernova data of Riess et al.\cite{REE} and
Perlmutter et al.\cite{PER}. The usual way to describe the structure
and evolution of our observable universe is to assume that on the
largest scales it is Friedmann-Lama\^ itre-Robertson-Walker
(FLRW-universe), i.e. isotropic and spatially homogeneous. The
observational parameters specifying the FLRW-universe are (i) the
Hubble parameter $H$ (ii) the deceleration parameter $q$ (iii) the
cosmological constant $\lambda$ (iv) the equation of state (v) the
ratio $\Omega_m$ of the matter density to the critical matter
density (vi) the spatial curvature. New observation suggests a
FLRW-universe that is light-weight ($\Omega_m<1$), is expanding
($H>0$), is cooling ($dT_U/dt<0)$), is {\it accelerating} ($q<0$),
and is flat ($\sum \Omega_i=1$), where $i$ denotes the number of
components occurring in the universe \cite{PER} \cite{pe} \cite{BO}.

To account for cosmic acceleration it is necessary to take into
consideration a new type of energy, the dark energy, a hypnotical
form of energy which permeates all space and tends to increase the
rate of expansion of the universe. It is usually modeled as the
static cosmological constant, an energy density filling space
homogeneously and the quintessence, a dynamical, spatially
inhomogeneous form of energy with negative pressure \cite{5}.

Dark matter is causative agent of the current accelerating
expansion. This agent (stuff) must have negative pressure, in order
to produce acceleration of the cosmic scale factor. The essential
properties of dark energy can be summarized in the following points:
(a) It does not show its presence in the galactic space; (b) it is
relatively smoothly distributed in cosmic space; (c) it does not
emit and absorb elm radiation; (d) it has large, negative pressure;
(e) it forms approximately homogenous stuff.

An adequate and coherent cosmological theory, conform with recent
observation, should give at least answers to
the following problems \cite{P}:\\
(i) {\it The nature of the vacuum energy.}  In the literature, the
vacuum energy is theoretically modeled by many ways, e.g. as (i) a
very small cosmological constant (e.g.\cite{4}) (ii) quintessence
(e.g.\cite{5}) (iii) Chaplygin gas (e.g.\cite{6}) (iv) tachyon field
(e.g.\cite{7} \cite{PA} \cite{PAD}) (v) interacting quintessence
(e.g.\cite{8}),  quaternionic field (e.g.\cite{IK}), etc. It is
unknown which of the said and the expected follow up models will
finally emerges as the successful one.\\
(ii){\it The cosmological constant problem.} The '$\Lambda$-problem'
can be expressed as discrepancies between the negligible value of
$\Lambda$ for the present universe and the value $10^{50}$ times
larger expected by Glashow-Salam-Weinberg model
or by GUT  where it should be $10^{107}$ times larger.\\
(iii) {\it The fine-tuning problem}. It is a puzzle why the
densities of dark matter and dark energy are nearly equal today when
they scale so differently during the expansion of the universe.
Assuming that the vacuum energy density is constant over time and
the matter density decreases as the universe expands it appears that
their ratio must be set to immense small value ($\approx 10^{-120}$)
in the early universe in order for the two densities to nearly {\it
coincide} today, some
billions years later.\\
(iv) {\it The flatness problem}. Inflation predicts a spatially flat
universe. According to Einstein's theory, the mean energy density
determines the spatial curvature of the universe. For a flat
universe, it must be equal to the critical energy. The observed
energy density is about one-third of critical density. The
discrepancy between the value of the observed energy
density and the critical energy represents the flatness problem.\\

Especially interesting is the problem of acceleration of the
expansion of the universe which is generally solve by assuming a
vacuum energy. According to Glimer \cite{GL} the vacuum energy must
satisfied the following requirements: (i) It should be intrinsically
relativistic quantity having the dimension of the energy density.
(ii) It should be smoothly distributed throughout the universe.
(iii) It should cause the speedup of the universe. (iv) It should
balances the total mean energy density to $\Omega=1$.

As is well-known the FLRW cosmological model constitutes the
standard paradigm of present day cosmology. Its 4-curvature is
determined from the various contributions to its total
energy-momentum tensor, mainly in form of matter energy, radiation
pressure and cosmological constant. The cosmological constant
contribution to the curvature of space-time is represented by the
$\lambda$ term which enters the gravitational field equations in the
form
$$R_{\mu\nu}-\frac{1}{2}g_{\mu\nu}R=8\pi G \tilde T_{\mu\nu},\quad\eqno(1)$$
where $\tilde T_{\mu\nu}=T_{\mu\nu}+g_{\mu\nu}\Lambda (t).$
$T_{\mu\nu}$ is the ordinary energy-momentum tensor associated to
isotropic matter and radiation. When modeling the expanding universe
as perfect fluid with velocity 4-vector field $U^{\mu}$, we have
$$T_{\mu\nu}=-pg_{\mu\nu}+(\rho +p)U^{\mu}U^{\nu},\quad\eqno(2)$$
where $p$ is the isotropic pressure and $\rho$ is the proper energy
density of matter. With the generalized energy-momentum tensor, and
in the FLRW metric (k=0 for flat, $k=\pm1$ for spatially curved
universe)
$$ds^2=dt^2-R^2(t)\left (\frac{dr}{1-kr^2}+r^2
d\theta^2+r^2sin^2{\theta}d\phi^2\right )\quad\eqno(3)$$ the
gravitational field equations turns out to be the
Friedman-Lama\^{\i}tre equation
$$H^2\equiv \left( \frac{\dot R}{R}\right )^2
=\frac{8\pi G}{3} (\rho +\lambda)-\frac{k}{R^2}\quad\eqno(4)$$ and
the dynamical field equation for the scale factor gets the form
$$\ddot R=-\frac{4\pi}{3}G(\rho+3p-2\lambda)R\quad\eqno(5)$$

The baryonic contribution to the total matter content is far smaller
than the total amount of matter detected by dynamical means, namely
$\Omega^0_b\approx 5$ per cent of critical density. The total amount
of matter detected by dynamical means is $\Omega^0_M\approx 30$ per
cent of critical density. Therefore the bulk of the matter content
must be in the form of unknown kind of cold (non-relativistic and
non-baryonic ) invisible component. Significant amount of hot
(relativistic) dark matter are excluded because it would not fit
with the models of structure formation. So the radiation part at
present boils down to an insignificant fraction of neutrinos plus an
even more negligible contribution of very soft photons entering at
the level of one ten-thousandth of critical density

On the other hand, the astrophysical measurements tracing the rate
of expansion of the university with high-z Type Ia supernovae
indicate that $\Omega^0_{\lambda} \approx 70$ per cent of critical
energy density of the universe is cosmological constant or another
dark energy candidate with a similar dynamical impact of the
evolution of the expansion of the universe. Specifically, the
cosmical constant values found from Type Ia supernovae at high z is:
$$\Lambda_0=\Omega^0_{\lambda}\rho^0_c\approx 6 h_0^2\times
10^{-47} GeV^4.$$ Independent from these supernovae measurements,
the CMB anisotropy, including the recent data from WMAP satellite,
lead to $\Omega_0=1.02\pm 0.02.$ As a first observation, it is
obvious that this result leaves little room for our universe to be
spatially curve. As a second observation, when combining this result
with the dynamically determined values of the matter density, the
complete energy bookkeeping leads us to the conclusion:the rest of
the present energy budget must be encoded in the parameter
$\Omega^0_{\lambda}$ .

Although the standard cosmological hot model describes successfully
many features of the evolution of the universe the problem of the
cosmic acceleration requires its revision in that one adds to it
some new phenomenological components such as static cosmological
constant or quintessence scalar field. Here the question arises
whether there exists an alternative cosmological model based on
Newtonian physics in form of the Maxwell-like equations in flat
space-time which can describe the cosmic evolution in a comparable
way as nowadays cosmological models without any implanted
phenomenological parameters. The aim of this account is to give a
simple exposition of the gravitation described by means of
Maxwell-like equations especially with respect to the negative
energy of the gravitational field. We show that Since the negative
energy is linked with the negative pressure we show that it might be
the agent that causes the acceleration of the rate of cosmic
expansion. For the description of the dynamical evolution of the
universe we use the simplest isotropic and homogenous cosmological
model in its Newtonian analogue. General relativity reduces to
Newtonian gravity locally but when spatial uniformity exists then
local structure is equivalent to global structure \cite{BA}. We
study the action of the negative pressure due to the negative energy
of gravitation field and show that it might be a further component
of the universe having similar properties as the dark energy. As the
initial condition of the beginning of the cosmic expansion we
consider the extremely comprised unstable mass object with huge
density of the negative gravitation energy. The dynamical
instability of this object causes the uniformly dispersion of its
matter with extremely large velocity. This we consider as the
beginning of big bang. Finally, we describe the further fate of the
expanding universe and show that the negative energy can be taken as
important agent in the evolution  of the universe.

\section{The problem of negative energy in the Maxwell-like gravitation}

Maxwell, at the end of a paper entitled {\it A dynamical theory of
electromagnetic field} added a brief {\it Note on the attraction of
gravitation}. There was suggested that the energy density of a
gravitational field might be $-(8\pi)^{-1}
 R^2$ where $R$ is the
'intensity' of gravitational field, but Maxwell rejected the idea of
negative energy of the gravitational field, and insisted that it
negative value must be balanced by existence of an unknown intrinsic
positive energy in masses \cite{MX}.  The same rejection  again the
negative energy of gravitational field showed  also Helmholtz
\cite{HE}. In electrostatics, one derives an expression for the
energy density of the electrostatic field by calculating the work
done in assembling a charge distribution from elements of charge
that are initially in a dispersed state. A similar situation in
classical Newton theory leads to a strange conclusion that the
energy density of gravitational field is necessarily {\it negative}.

Recently, motivated by these facts, there has been much interest
paid to the consistent field equations for the gravitational
interaction in an analogy to the extended Maxwell elm theory. In
this approach the interaction energy of a mass body occurs,
similarly as in elm, in its neighboring space and is necessarily
negative. The quantitative characteristics of this field obey the
Maxwell equations. As is well-known the standard Maxwell equations
applied to gravity have the form \cite{IK}
$$\nabla \times \vec B -{\partial \vec E \over \partial ct} = \frac{4\pi}{c}\vec i
\quad\eqno(6a)$$
$$ \nabla \times \vec E +
{\partial \vec B \over \partial ct} =0 \quad\eqno(6b)$$
$$\nabla \vec E =\varrho,
 \quad \eqno(6c)$$
where $\vec E$ and $\vec B$ represent the familiar vector field
variables with the difference that the source of gravitational field
$\rho$ and $\vec i$ are purely imaginary quantity (We refer to
article of Ulrych \cite{UL} where the relevant references concerning
the Maxwell-like gravity theories are presented.) It has been shown
that the simplest way how to cross over from elm to gravitation
consists either in substituting for the electrical charge $e$ the
imaginary 'gravitational charge' $i\sqrt{G}M$ $(i=\sqrt{-1})$
\cite{IKK} or in taking the elm equations with the negative
permittivity $\epsilon =-1$ \cite{BR}. In both cases all measurable
quantities of gravitational interaction are real and obey
Maxwell-like equations but necessarily with opposite sign. For
example, in gravitostatics, Gauss's law in spherically symmetrical
case turns out to be
$$\bigtriangledown E_g=4\pi i \sqrt{G}\rho(r)\quad \eqno(6d),$$
whose solution is
$$E_g=\frac{4\pi \sqrt{G} i}{r^2}\int{r^2 \rho(r)} dr.$$
For the sake of simplicity, we illustrate this situation only for an
idealized model case with $\rho=q=const.$ The solution of Eq.(6d)
has the simple form
$$E_g=\frac{4\pi i}{3}\sqrt{G}qr$$
and the corresponding field energy density becomes
$$E_f=\frac{1}{8\pi} (E_g)^2.\quad\eqno(7)$$
The  total negative field energy of within the sphere of radius $R$
and mass $M$ assumes the form
$$U_g =\int_0^R E_f^2 4\pi r^2 dr=-kGq^2R^5= -\kappa
\frac{GM^2}{R},\quad\eqno(8)$$ where  $k=1.27$ and $\kappa=0.1.$ The
 energy content in this sphere is according to
Eq.(7) negative while that of $M$ is positive.  The sum of the
positive and negative energy containing in mass object is
$$ U_t=Mc^2+U_g=Mc^2\left(1-\kappa\frac{\lambda}{R} \right) \quad \lambda=\frac{GM}{c^2}.$$
If we denote the effective mass of $M$ by symbol $M^{(eff)},$ where
$M^{(eff)}=U_t/c^2$, then the force at the surface of this mass
object acting on test mass $m$ is given as
$$F=-\frac{GM^{(eff)}m}{r^2}=
-\frac{GMm}{r^2}\left (1-\kappa\frac{\lambda}{R}\right
).\quad\eqno(9)$$ We see that the force law (9) changes its sign
when $1-\kappa\frac{\lambda}{R}<0.$ If $U_t=0$ then the radius  of
this spherical body is $R_0=\kappa R_S=0.05 R_S,$ where $R_S$ is the
Schwarzschild radius. In this case, a spherical body with constant
density of radius $R_0$ has its total energy just equal to zero and
no gravitational field outside of it exists. The total energy of a
spherically symmetrical mass body can become even negative. This
happens when $R<R_0$. In that case the mass body becomes unstable.
The thee main consequences of the field approach to gravity are:\\
(i) The gravitational mass of gravitating body depends generally
on its shape.\\
(ii) If taking into account that the (negative) energy density of
the gravitational field is a source of the gravitational field
itself, then one obtains a slightly modified force law for the
gravitationally interacted bodies. In the simplest case of a
point-like gravitational source this leads to the equation
$$F'(r)+\frac{2}{r}F'(r)=-\frac{G^2M^2}{c^2r^4}\quad\eqno(10)$$
whose solution is
$$F'(r)=-\frac{GM^2}{r^2}+\frac{G^2M^2}{c^2r^3}. $$ We see that to
Newton's law is added a further term which modifies the force law
between gravitationally interacting bodies.\\ (iii) The negative
energy represents a kind of anti-gravitational force similar as the
dark energy in standard cosmology. \\

\section{The negative energy in the cosmology-the Newton-like cosmological model.}

The fully relativistic Friedman equation for homogeneous and
isotropic universe has the form
$$\frac{1}{R^2}\left (\frac{dR}{dt}\right )^2 =\frac{8\pi
G\rho}{3}-\frac{k}{R^2},\quad\eqno(11a)$$ where $\rho$ is the
density of matter and $k=\pm 1$ or $0$. As first noted Milne and
McCrea in their classical article \cite{MM}, the Newtonian equation
governing the evolution of a particle with mass $m$ and total energy
$E$ located a distance $R$ from the center of a homogeneous and
isotropic sphere of matter with density $\rho(t)$ which expresses
the conservation of energy is
$$\frac{1}{R^2}\left (\frac{dR}{dt}\right )^2= \frac{8\pi G \rho}{3}-\frac{(-2E/m)}{R^2}.\quad\eqno(11b)$$
If we set $(-E/m)=k$, then equations (11a) and (11b) are identical.

Instead of deriving Eq.(11b)  using the energy conservation, we
start with the familiar Newtonian force law. One picks an arbitrary
point in space as the origin of coordinates, and considers the
gravitational force acting on a test body $m$ a distance $R$ from
this center . The Newtonian force law leads to the rule that for a
spherically symmetrical mass distribution, only the mass inside the
sphere of radius $R$ has a net gravitational effect, and so the
force acting on $m$ is
$$m\ddot R=-\frac{GMm}{R^2},$$
or using $M=(4 \pi/3) \rho R^3,$
$$\ddot R =-\frac{4\pi \rho}{3}R.\quad\eqno(12)$$
Assuming that the density varies $R$ as $\rho=\rho_0R_0^3/R^3$, then
by inserting this expressing for $\rho$  into Eq.(12) and
multiplying by $\dot R$, the the results can be integrated to obtain
an equation identical to Eq. (11b),but with a different integration
constant.

 We consider the universe as a sphere of the radius $R_u$ in the flat space filled
by homogenous and isotropic matter governed by  the force law (9).
Such a cosmological model we shall call as the Newton-like universe.
Its total matter content $M_u$ does not vary with time while its
total negative mass content amounts
$${\bf M_g}=-\frac{GM_u^2}{c^2R_u}=-\frac{M_uL_u}{R_u}\qquad
L_u=\frac{\kappa GM_u}{c^2}.$$ This follows from Gauss's law for a
spherical body $M$ of radius $R$ with isotropic and homogeneous
matter density. Accordingly, the total energy of the universe is
given as
$${\bf E_t}=M_uc^2-\frac{\kappa GM_u^2}{R_u}$$
and the force acting on mass $m$ located a distant $R$ from the
center of sphere is
$${\bf F}=m\ddot R=-mG\left (\frac{M}{R^2}-\frac{\kappa GM^2}{c^2R^3}\right )
=-\frac{mGM}{R^2}\left (1-\frac{L_u}{R}\right ),\quad\eqno(13)$$
where $M=(4\pi /3) \rho R^3,$ $\rho$ being the matter density in
this sphere. First integral of Eq.(13), expressing the conservation
of energy, gets the form
$$\frac{1}{2}(\dot R)^2=\frac{GM}{R}-\frac{G^2M^2}{2c^2R^2}-\left(\frac{-E}{m}\right )=
\frac{GM}{R}-\frac{GML}{2R^2}-\left (\frac{-E}{m}\right).$$
Accordingly, we have
$$\dot R=\sqrt{2(T_1(R)+T_2(R)+T_3)},$$
where
$$ T_1(R)=\frac{GM}{R},\qquad T_2(R)=-\frac{\kappa G^2M^2}{c^2R^2}=\frac{GML}{R^2}\qquad
{\rm and}\qquad  T_3=-\left (\frac{-E}{m}\right ).$$ The first term
$T_1$ represents the gravitational {\it attraction}, the second term
$T_2$ the gravitational {\it repulsion} and the third term $T_3$ the
total energy divided by $m$. According to the relations among $T_1$,
$T_2$ and $T_3$ the velocity $\dot R$ is decreasing or increasing.
The velocity  $\dot R$ assumes maximal value for $R_{max}=\kappa
R_S/2,$ where $R_S$ is the Schwarzschild radius of matter containing
in the sphere of radius $R$.

The scenario of the proposed Newton-like cosmological model can be
briefly sketched as follows.  At the big bang, the whole matter
content of the universe was extremely  comprised in a sphere of
radius $R_0\ll R_S $ so that the negative energy considerable
prevailed the positive energy of $M_uc^2$. ( $R_0$ and $M_0$ we take
the initial conditions for the beginning of cosmic evolution). The
large repulsive force acting on the individual spherical mass shells
of $M_u$  depends, according to Eq.(13), on $R$. The largest force
acted on the surface mass shell. As a consequence of this force the
matter was uniformly spread outwards starting a kind of cosmic
inflation. The acceleration at the beginning of cosmic evolution was
extremely large and then, in course of time, it became zero. This
happened when $R=GM/(2c^2)$. Then the acceleration switched to
deceleration.

Given $\rho$, Eq.(13) can be re-written as
$$\ddot R=-\frac{4\pi}{3}G\rho R+(\frac{4\pi}{3})^2\frac{G^2\rho^2 R^3}{c^2}=
-(\frac{4\pi}{3})GR\rho \left (1-(\frac{4\pi}{3})\frac{G\rho
R^2}{c^2}\right ).\quad\eqno(14)$$
or
$$\ddot R=-\frac{4\pi}{3}G\left (\rho-\rho L\right )R,\quad\eqno(15) $$
where
$$L(\rho,R)=\frac{4\pi G\rho R^2}{3c^2}.$$
$L$ is a dimensionless quantity. If one takes a point in the
Newton-like universe as the origin of coordinates, then for a
spherically symmetrical mass distribution only the mass equivalent
to total energy inside the radius $R$ has a net gravitational
effect, therefore $\ddot R$ is negative for $L<1$. For example, the
mean mass density in the present day universe $\rho \approx 10^{-29}
gr.cm^{-3}$. When taking into account that $G/c^2\approx 10^{-28}
cm.gr^{-1}$, the repulsive prevails the attractive force at the
distance $R_a\approx 10^{27}-10^{28} cm$. This distance
approximately corresponds the distance from where the rate of cosmic
expansion begins to be accelerated. For $R\ll R_a$, the first term
$GM/R$ in Eq.(13) becomes larger than $G^2M^2/(c^2R)$ so that in the
sphere of this radius the rate of expansion is similar as in
standard cosmological model.

The formula for $\ddot R$  in the Friedman cosmology for matter
dominated era is
$$\ddot R =-\frac{4\pi G}{3}(\rho_M-2\Lambda)R\quad\eqno(16)$$
The comparison of Eq.(16) and Eq.(15) yields
$$\Lambda=\frac{L\rho}{2},$$
i.e. $\Lambda$ is proportional to $\rho$. However, $\Lambda$ does
not represent the static cosmological constant because it is a
function of $R$.

The well-known Whitrow-Randal relation \cite{W} \cite{R} one derives
putting $U_t=0$ which yields
$$\frac{\kappa GM_u^2}{R_u}\approx M_uc^2$$
Taking $M\approx 10^{56}-10^{58} g$ and $R=10^{26}-10^{28}cm$ we find that
$$R_u\approx \frac{GM_u}{c^2}\approx \frac{R_S}{2},\quad\eqno(a)$$
where $R_S$ is the Schwarzschild radius of the Universe. Eq.(a)
expresses that at present time the negative gravitational energy is
approximately equal to its total positive energy.

The Newton-like field approach to gravity can be applied also to
astrophysics. It is generally accepted that when a star run out of
nuclear fuel, the only force left to sustain it against gravity is
the pressure associated with the zero-point oscillation of its
constituent fermions. This is valid if the gravitational force obeys
Newton's law. If one takes, instead of Newton's, the Newton-like
force law (13) then the force at the surface of a star depends on
its total (negative and positive) energy. When the star is shrinking
to a sufficiently small volume the gravity at its surface is
weakening until it complectly ceases. The weakening of force near to
the gravity center in the Newton-like field approach  makes possible
the existence of star-like objects in equilibrium which are more
collapsed than neutron stars (see \cite{GM}).

As is well-known general relativity offers the possibility of stable
end state of collapsing mass body called black hole representing a
mass object from which elm radiation can not escape. The black hole
in its general form is conditioned with the {\it cosmic censorship}
hypothesis which is not proved so far \cite{PE}. This censorship
hypothesis represents  the major unsolved problem of classical
general relativity today \cite{PEN}. The question, whether an event
horizon will be formed around any singularity to screen it from the
outside world is until now not answered. If cosmic censorship were
not true, there see to be nothing in general relativity (and in
Newton's gravity) to prevent that a star can finally come to rest
only in a configuration of zero volume. In this process an infinity
amount of gravitational energy were pumped into kinetic energy of
star's matter as well as into its thermal energy. In absent of a
protective event horizon, all of this energy will be released to
outside world. It is often argued that a certain confirmation for
the existence of cosmic censorship  for the general non-symmetrical
collapse is the fact that such an event would be with certainty
notified by astronomers during the whole period of the astronomical
observation. However, this argument can be reversed, i.e. such an
event (the unlimited release of energy by mass collapse) has not
been observed because the end state of gravitational collapse is not
an object screened by horizon of event but an object smaller then
neutron star, the energy release of which is at its collapse
limited.

Interestingly, the gravitostatics can be straightforwardly extended
to a complete gravitation Maxwell-like field theory describing by a
whole set of the Maxwell-like equations - gravitodynamics.  Singh
(see \cite{UL}) has shown that the tree main post-Newtonian solar
solar system experiments can be also explained in the frame of
gravitodynamics.

\section{Conclutions}

(i) The Maxwell-like field approach to gravity leads necessarily to
the existence of the negative gravitational field energy.\\
(ii) The gravitational mass of gravitating bodies is given by the
total energy containing in them.\\
(iii) Application of the Maxwell-like gravity field equations to
cosmology leads to the Maxwell-like universe described essentially
by two components: the positive matter energy and the negative
gravitation field energy.\\
(iv) The cosmic evolution began by an explosion-like uniform
dispersion of a initial very comprised piece of the spherically
symmetrical matter $M_0$ with the radius $R_0\ll R_S,$ where $R_S$
denotes the Schwarzschild radius assigned to $M_0$. This process
resembles the cosmic inflation. \\
(v) The gravitational negative field energy  exhibits some similar
properties as dark matter, namely it has negative pressure and it is
smoothly distributed  in cosmic space.\\
(vi) The 'cosmological term' $\Lambda=(\rho L)/2$ is proportional to
$\rho$  which might resoled the fine tuning problem \cite{RN} \cite{AR}.\\
(vii) The negative energy of gravitational field within the
star-like object causes that its shrinking leads, in the idealized
case, not to its zero volume but to its stable end state having
approximately the radius equal to its Schwarzschild radius.\\
(v) The initial conditions for cosmic evolution are the total mass
$M_u$ and $R_0$. The subsequent cosmic evolution is essentially
given by these two quantities.

The aim of this article was only to outline the basic ideas
concerning the Maxwell-like field theory of gravitation and its
application to cosmology. This is why everything is simplified and
many important issues remained open.


\begin{thebibliography}{99}
\bibitem{TO}
J. L. Tonry et al., ApJ {\bf 594} (2003), 1.
\bibitem{BAA}
B. J. Barris et al., ApJ {\bf 602} (2004), 571.
\bibitem{RS}
A. G. Reiss et al.,ApJ {\bf 607} (2004), 665.
\bibitem{MX}
J. M. Maxwell, The Scientific Papers of James Clark Maxwell, Vol. 1
(New York, 1965), p.571.
\bibitem{IKK}
V. Majern\'{\i}k, Astrophys. Space Sci. {\bf 14} (1971), 265.
\bibitem{BR}
L.Brillouin, Relativity Reexamined. Academic Press, New York, 1970.
\bibitem{UL}
S. Ulrych, Phys.Lett. {\bf B 633} (2006), 631.
\bibitem{PE}
R. Penrose, Rivista del Nuovo Cimento {\bf 1} (1969), 252.
\bibitem{PEN}
R. Penrose, In:{\it Theoretical Principles in Astrophysics and
Relativity} (ed. by N.R. Lebowitz, W.H. Reid and P. O. Vandervoort).
 University of Chicago Press 1978, p.217.
\bibitem{HE}
E. Whittaker, History of the theories of aether and electricity: the
classical theories. Nelson, London, 1951, pp. 217-218.
\bibitem{MM}
 E. A. Milne and McCrea, Q. Jl. Math. (Oxford) {\bf 5} (1934), 73.
\bibitem{BA}
J. D. Barow, In:{\it Gravitation in Astrophysics}, Carg\`{e}se
Lectures, pp. 239-306, eds B. Carter and J.Hartle, Plenum , New
Yourk 1987.
\bibitem{W}
G. Whitrow, Nature {\bf 158} (1951), 165.
\bibitem{R}
G. Whitrow and D. Randall, Mon. Not. R. Astron. Soc. {\bf 111}
(1951),455.
\bibitem{PER}
S. Perlmutter et al. ({\it the Supernova Cosmological Project})
  Astroph. J. {\bf 517} (1999), 565.
\bibitem{REE}
A. G. Reiss et al. ({\it the High-z SN Team}) Astronom. J. {\bf 116}
(1998) 1009.
\bibitem{pe}
P, Peebles, Nature (London) {\bf 398}, (1999), 25.
\bibitem{BO}
N. A. Bahcall, J. S. Ostriker, S. Perlmutter and P. Steinhardt, {\it
Science} {\bf 284}, (1999), 1481.
\bibitem{T}
J. L. Tonry et al., Astrophys. J. {\bf 594}, (2002), 1
\bibitem{MK}
V. Majern\'{\i}k, Gen. Relat. Grav. {\bf 35}, (2003), 1831.
\bibitem{2}
2. Proc. of I.A.P. Conference "On the Nature of DE held in Paris
(1-5 July 2002) edited by P. Brax, J. Martinn and J. P. Anduzan
(Frontier Group, Paris, 2002).
\bibitem{4}
A. Melchiorri, The death of quintessence?, In Proc. of the I.A. P.
Conference " On the Nature of DE" held Paris,(1-5 July 2002), edit.
P. Brax, J. Martin and J.P. Uzan (Frontier Group, Paris, 2002.
\bibitem{5}
C. R. R. Calwell, R. Dave, and Steinhardt, Phys. Rev. Lett. {\bf
80}, (1998), 1582.
\bibitem{6}
A. Kamenshichik et al, Phys. Lett. B {\bf 511} (2001), 265.
\bibitem{7}
J. S. Bagla et al, Phys Rev D {\bf 67} (2003), 063504.
\bibitem{8}
L. P. Chimento et al Phys. Rev. {\bf D 67} (2003), 063504.
\bibitem{IK}
V. Majern\'{\i}k, {\it Gen. Relat. Grav.} {\bf 36} (2004), 2139.
\bibitem{GL}
F. Glimer, JETF {\bf 49}, (1965), 542.
\bibitem{GM}
V. Majern\'{\i}k and J. Gembarovi\v c, Astrophys. Space sci. {\bf
32} (1990), 250 .
\bibitem{P}
F. Piazza and Ch. Marinoni, Phys. Rev. Lett. {\bf 91} (2003) 141301.
\bibitem{PA}
T. Padmanabdan, {\it Phys. Rev. D} {\bf 66} (2002),021301
(hep-th/0204120).
\bibitem{PAD}
T. Padmanabdan, T. Roy Choudhury, {\it Phys. Rev. D} {\bf 66}
(2002), 081301 (hep-th/0204120).
\bibitem{RN}
V. Majern\'{\i}k, Phys. Lett. {\bf A 282} (2001), 362.
\bibitem{AR}
A.-M., M. Abdel-Rahman and Ihab F. Riad, Astron. J. {\bf 134}
(2007), 1391.
\end{thebibliography}
\end{document}